\documentclass{ifacconf}  % Comment this line out if you need a4paper

\usepackage{graphics} % for pdf, bitmapped graphics files
\usepackage{epsfig} % for postscript graphics files
\usepackage{amsmath} % assumes amsmath package installed
\usepackage{amssymb}  % assumes amsmath package installed
\usepackage{booktabs}
\usepackage{csquotes,comment}
\usepackage{multirow}
\usepackage{xcolor}
\usepackage{natbib}        % required for bibliography

\newtheorem{define}{Definition}

\begin{document}
\begin{frontmatter}
\vspace{-1em}
\title{Learning myopic mixed-integer nonlinear model predictive control from expert demonstrations} 

\author[Tue]{Christopher A. Orrico} 
\author[Tue]{W.P.M.H. Heemels} 
\author[Tue,NTNU]{Dinesh Krishnamoorthy}

\address[Tue]{Department of Mechanical Engineering, Eindhoven University of Technology, 5600 MB Eindhoven, The Netherlands. (e-mail: c.a.orrico@tue.nl, m.heemels@tue.nl).}
\address[NTNU]{Department of Engineering Cybernetics, Norwegian University of Science and Technology, O. S. Bragstads Plass 2D, 7034 Trondheim, Norway (e-mail: dinesh.krishnamoorthy@ntnu.no)}

\begin{abstract}                % Abstract of 50--100 words
Applying nonlinear model predictive control (NMPC) to systems with hybrid dynamics or discrete actions typically yields mixed-integer nonlinear programs (MINLPs), whose real-time solution remains a major challenge and limits the applicability of mixed-integer NMPC (MINMPC). This paper proposes a myopic MINMPC framework that incorporates value-function approximation to substantially reduce the online computational burden. Using Bellman’s principle of optimality, we shorten the prediction horizon and append a value function learned offline from expert \emph{state–action} demonstrations via inverse optimization with optimality residual minimization. A central feature is the dual treatment of discrete decisions, whereby integer constraints are relaxed during offline learning to enable KKT-residual-based value function synthesis, while the online controller enforces the true integer constraints to ensure feasibility. The learned value function induces a policy that is approximately \emph{policy-consistent} with the expert demonstrations. The resulting controller achieves high closed-loop performance with a significantly shorter horizon, enabling real-time MINMPC. The effectiveness of the approach is demonstrated on the Lotka–Volterra fishing problem and a satellite attitude control system with discrete actuators.
\end{abstract}


\end{frontmatter}
%%%%%%%%%%%%%%%%%%%%%%%%%%%%%%%%%%%%%%%%%%%%%%%%%%%%%%%%%%%%%%%%%%%%%%%%%%%%%%%%

\section{Introduction}
\label{sec:introduction}
\vspace{-1em}
Model predictive control (MPC) is a powerful tool for controlling complex dynamical systems that arise in many applications. MPC formulates the control problem as a discrete-time, finite-horizon optimal control problem (OCP). At each time step in the control loop, the OCP is solved and the first control input is implemented. In application domains with discrete actuators, switching systems, disjoint constraint sets,  specified logics, etc., MPC may be applied by formulating the OCP with both continuous and integer decision variables \citep{richards2005MIP-MPC,bemporad1999control}. The resulting mixed-integer nonlinear MPC (MINMPC) framework requires the online solution of mixed-integer nonlinear programming (MINLP) problems at each sampling instant. MINLP problems are generally $\mathcal{N}\!\mathcal{P}$-hard, making them challenging for real-time control \citep{bemporad1999control}.

This is especially problematic when controllers must consider the effect of the control actions far into the future (long prediction horizons) to achieve better performance. The size of the optimization problem and, thus, computational cost, increases with the length of the prediction horizon. This is further amplified when solving MINLPs, where computational cost scales exponentially with the number of decision variables \citep{bemporad1999control}. 

\vspace{-0.25cm}
\subsubsection*{Related work} 
Various strategies and heuristics have been proposed to address real-time MINMPC problems. These include solving the relaxed problem by removing integer constraints and obtaining the integer trajectory through rounding schemes like sum-up-rounding \citep{sager2009reformulations} or the penalty term homotopy method \citep{sager2006numerical,orrico2023mixed}. Piecewise linearization may be used to transform the MINLP into mixed-integer linear or quadratic programs. However, such approximations may not always be applicable or sufficiently accurate. Real-time iteration schemes for MINMPC based on outer convexification and rounding have been extended \citep{deMauri2020real}, although this method is challenging if discrete variables appear in inequality constraints. For switched linear systems, \cite{antunes2016linear} use existing knowledge of policy bounds or stabilizing policies to inform policy estimation. In other direct policy approximation approaches, the MPC policy is approximated using deep neural networks trained on large amounts of offline-generated state-action data pairs \citep{karg2018deep}. This requires extensive data covering the entire feasible state space and poses challenges for any later updates in the controller \citep{krishnamoorthy2021sensitivity,mesbah2022fusion}. The burden of generating the training data set only increases further for MINMPC. 

Seemingly, a straightforward approach would be to shorten the online prediction horizon, reducing the problem size and, consequently, the computational cost. Naively using a short prediction horizon can lead to performance degradation. We can, however, leverage Bellman's principle of optimality to split the problem into two parts, where the optimal value function of the tail part of the prediction horizon is replaced by a value function that captures the remaining \emph{cost-to-go}. The online controller is then reduced to as small as a one-step-look-ahead controller without jeopardizing performance \citep{bertsekas2019reinforcement}. Computing the true value function is an intractable problem, so instead approximate value functions can be employed. This has been predominantly studied under the context of approximate dynamic programming (ADP) and reinforcement learning (RL), whereby the value function is learned from closed-loop performance data, cf. \cite{bertsekas2019reinforcement, powell2007approximate, mesbah2022fusion} and the references therein. Generally, learning from closed-loop performance data (as done in RL) requires  several episodes of interaction with a real system that trade-off exploration and exploitation to learn the value function. The ability to interact with the real system (or a high fidelity simulation of the system) may not always be possible in many engineering applications. This motivates the use of offline expert demonstrations to learn a suitable value function.  

Some recent works have explored learning value functions from expert demonstrations in the form of optimal state–value pairs \citep{abdufattokhov2021learning,DK2025c2gcdc,Chatzikiriakos2024learningV}. Within the  context of learning from demonstrations (LfD), however, expert data typically consists of optimal state–action pairs rather than state–value pairs, especially if given by alternative control policies, or human experts. While standard supervised learning methods based on empirical loss minimization are a default choice when optimal state-value pairs are available, learning a value function directly from optimal state–action demonstrations requires inverse optimization, which seeks to recover the parameters of an optimal control problem that rationalize an observed policy \citep{ab2020inverse}. Prior work has applied inverse optimization to linear MPC \citep{keshavarz2011imputing, akhtar2021learning} and nonlinear MPC (NMPC) \citep{molloy2018finite}. 
A particularly promising technique in this context is the Karush–Kuhn–Tucker (KKT) residual minimization \citep{keshavarz2011imputing}, which learns a value function whose parameters make the demonstrated state–action pairs approximately satisfy the KKT conditions. This differs fundamentally from the empirical loss–based approaches \citep{ab2020inverse}. However, extending KKT residual minimization to mixed-integer MPC is not straightforward, as classical KKT conditions apply only to continuous optimization problems and do not directly characterize optimality in mixed-integer programs. This constitutes the first major bottleneck addressed in this paper, namely finding a way to leverage KKT-based inverse learning in the MINMPC context. A second challenge is the prohibitive cost of generating expert demonstrations, which requires solving numerous long-horizon mixed-integer OCPs offline. Our work aims to also address this challenge.

\vspace{-0.5cm}
\paragraph*{Main contribution} This paper extends inverse optimization via KKT residual minimization to the setting of NMPC with mixed-integer decision variables, thereby enabling short-horizon MINMPC. The central idea is to relax the integer constraints during the offline learning phase. This relaxation makes it possible to formulate the classical KKT conditions and apply optimality-residual minimization in a context where such conditions would otherwise not exist for the original MINLP. In addition, we show that the offline state–action data can also be generated from continuous relaxations of the MINMPC problem. The resulting approximate cost-to-go function learned from  state-action demonstrations is then used in an online myopic MINMPC controller. This architecture simultaneously (i) overcomes the theoretical barrier posed by the absence of KKT conditions in mixed-integer programs, and (ii) alleviates the computational difficulty of generating long-horizon mixed-integer demonstrations, since the relaxed learning problems are substantially cheaper to solve. 

Together, these elements form a practical and scalable approach for MINMPC with learned value function approximations. Our proposed strategy is based on the observation that MPC-based schemes can still exhibit good closed-loop performance even when the learned cost-to-go function is only approximate \citep{wang2015approximate}. Unlike policy learning from expert demonstrations, exact recovery of the expert’s value function is not necessary. The approximate value function merely needs to be informative enough to compensate for the shortened horizon of the resulting myopic MPC formulation.

\section{MINMPC problem formulation}
\label{sec:problemFormulation}
%\subsection{MINMPC Problem}\label{subsec:MINMPC}
\vspace{-0.1cm}
Consider the MINMPC problem
\begin{subequations}\label{Eq:MINMPC}
	\begin{align}
		V^*(x(t)) = \min_{\substack{x_0, \ldots, x_N \\ w_0, \ldots, w_{N-1}  }} \; &\sum_{k=0}^{N-1}\ell(x_{k},w_{k}) + V_f(x_N), \label{Eq:MINMPCobj}\\
		\text{s.t.} \; \; x_{k+1} = f(x_{k},w_{k}), \; \;  &k = 0,1,\dots,N-1, \label{Eq:MINMPCf}\\
		  g(x_{k},w_{k}) \leq 0, \; \; & k = 0,1,\dots,N-1, \\
         x_0 = x(t), \; \;&\\
         x_N \in \mathcal{X}_f, \; \; &
	%	& \qquad \forall k = 0,\dots,N-1 \nonumber
	\end{align}
\end{subequations}
with state $ x_k \in \mathbb{R}^{n_{x}} $ and decision variables $ w_k:= [u_k,z_k]^{\mathsf{T}} $. Here, $u_k \in \mathbb{R}^{n_{u}} $ and $ z_k \in \mathbb{Z}^{n_{z}} $ denote the set of continuous and discrete controls, respectively, predicted over the horizon of length $ N \in \mathbb{N}$ from an initial state $x(t)$ at discrete-time step $t$. The stage cost is defined as $ \ell:  \mathbb{R}^{n_{x}} \times \mathbb{R}^{n_{u}} \times \mathbb{Z}^{n_{z}}  \rightarrow \mathbb{R}$, system dynamics are $ f: \mathbb{R}^{n_{x}} \times \mathbb{R}^{n_{u}} \times \mathbb{Z}^{n_{z}}  \rightarrow \mathbb{R}^{n_{x}} $, and the  constraints are $g:\mathbb{R}^{n_{x}} \times \mathbb{R}^{n_{u}} \times \mathbb{Z}^{n_{z}}  \rightarrow \mathbb{R}^{n_{g}}$. At each discrete-time step, the first control element $w_{0}^* =[u_{0}^*,z_{0}^*]^{\mathsf{T}}$ of an optimal MINLP solution $(x_{0}^*,\ldots,{x_{N}^*},w_{0}^*,\ldots,{w_{N-1}^*})$ \eqref{Eq:MINMPC} is implemented and the above process is repeated at $t+1$, resulting in a control policy $w^*(t) = \pi(x(t))$. 

%\begin{assume}
%$N$ in \eqref{Eq:MINMPC} is chosen to be sufficiently large such that  $\ell(x_{k},w_{k}) = 0  ~ \forall ~ k > N$, $\forall \; x(t) \in \mathcal{X}_{feas}$, where $\mathcal{X}_{feas}$ is the feasible state-space. 
%\end{assume}

\textit{Problem setting:} The MINMPC problem \eqref{Eq:MINMPC} is well designed with a sufficiently long horizon length $N$, such that the terminal set $\mathcal{X}_f$ does not restrict the feasible state-space $\mathcal{X}$. However, the MINMPC is computationally too intensive to solve within the desired computation time, thereby prohibiting real-time implementation.

\textit{Objective:} Learning from demonstrations of the \enquote{expert} MINMPC \eqref{Eq:MINMPC} queried offline, we aim to build a simpler control policy that is amenable for online implementation.

The OCP for the tail of the prediction horizon starting at the state $x_{\eta}$ at prediction step $\eta \ll N$ can be denoted by
\begin{subequations}
	\begin{align}
		V_{\eta}(x_\eta) = \min_{\substack{x_\eta, \ldots, x_N \\ w_\eta, \ldots, w_{N-1}  }} \; & \sum_{k=\eta}^{N-1}\ell(x_{k},w_{k}) + V_f(x_N) \\
		\text{s.t.} \; \; x_{k+1}  = f(x_{k},w_{k}), \; \;  &k = \eta,\eta+1,\dots,N-1,   \\
		  g(x_{k},w_{k}) \leq 0, \; \;  &k = \eta,\eta+1,\dots,N-1 \\
    x_N \in \mathcal{X}_f, \; \; &
	\end{align} \label{eq:DP}%
\end{subequations}
where $ V_{\eta} $ is the optimal-cost-to-go when starting from state $ x_{\eta}$. By Bellman's principle of optimality, we can truncate  \eqref{Eq:MINMPC} to equivalently solve
\begin{subequations}\label{Eq:DPMINMPC}
\begin{align}
	 V^*(x(t)) = \min_{\substack{x_0, \ldots, x_\eta \\ u_0, \ldots, u_{\eta-1}  }} \; &\sum_{k=0}^{\eta-1}\ell(x_{k},w_{k}) + V_{\eta}(x_{\eta}),  \\
	\text{s.t.} \; \; x_{k+1}  = f(x_{k},w_{k}), \; \;& k = 0,1,\dots,\eta-1, \\
     g(x_k,w_k) \leq 0, \; \; & k = 0,1,\dots,\eta-1,\\
     x_0=x(t), \; \; &
\end{align}
\end{subequations}
given the \textit{true} cost-to-go $V_{\eta}(x)$ appended to the truncated problem. In separating the value function in this way, we significantly reduce the online complexity of the MINMPC controller, which now only must solve for decisions $w_k$ for $k = 0,\dots,\eta-1$, where $\eta$ can be as small as 1.

\section{Learning the cost-to-go with KKT residual minimization}
Unfortunately, calculating the optimal value function $V_{\eta}(x)$ in \eqref{Eq:DPMINMPC} for all possible states is an intractable problem. Instead, we propose using inverse optimization to synthesize a function $\hat{\mathcal{V}}(x,\theta)$, parameterized by $\theta$, from expert demonstrations of \eqref{Eq:MINMPC}. Here, we take the extreme case of $\eta = 1$ to achieve the largest possible reduction in controller complexity over \eqref{Eq:MINMPC}, although any $\eta \ll N$ could be used in principle. We formulate a myopic MINMPC problem from \eqref{Eq:DPMINMPC} with $\hat{\mathcal{V}}(x,\theta)$ for $\eta = 1$, and  some  $x(t)$ as 
\vspace{-0.5em}\begin{subequations}\label{Eq:ADPMINMPC}
\begin{align}
	\min_{w} \; & \ell(x(t),w) + \hat{\mathcal{V}}(f(x(t),w), \theta),  \\
	\text{s.t.} \; & g(x(t),w)\leq 0.
\end{align}
\end{subequations}
\begin{define}[Expert demonstration]
The expert policy $ \pi(x_{i}) = w_{i}^*$ is observed by querying the expert for different states $ x_{i} \in \mathcal{X}, ~i = 1,2,\ldots,M$ (which need not be ordered in time), each generating an optimal state-action pair $(x_{i},w^*_{i})$. This yields the data set $ \mathcal{D}:= \{(x_{i},w_{i}^*)\}_{i=1}^M $. 
\end{define}
While we consider the setting where \eqref{Eq:MINMPC} is the expert, expert demonstrations could come from any suitable policy, including human experts. We assume that the expert demonstrations are primal feasible (i.e., $g(x_{i},w^*_{i}) \leq 0,~i = 1,2,\ldots,M$). 
%To compute $\hat{\mathcal{V}}(x,\theta)$, we first generate a set of optimal state-action pairs $ \mathcal{D}:= \{(x_{i},w_{i}^*)\}_{i=1}^M $ by solving the long prediction horizon MINMPC \eqref{Eq:MINMPC} offline (i.e., the expert demonstrations). 
Our aim is to compute $\hat{\mathcal{V}}(x,\theta)$ from the expert demonstrations $ \mathcal{D}:= \{(x_{i},w_{i}^*)\}_{i=1}^M $ such that we obtain a policy that is consistent with the expert demonstrations. 
In classical statistical estimation, consistency refers to the property that an estimator $\hat{\theta}$ converges in probability to the true parameter $\theta^*$ as $M \rightarrow \infty$.
In our setting, however, the primary objective is not \textit{exact recovery} of the value function, but rather \textit{recovery of the policy induced by it} when used as the terminal value function in the myopic MPC formulation.
This motivates the following notion of policy consistency.
\begin{define}[Policy consistency]
    The control policy $\hat{\pi}_\theta(x)$ obtained by solving \eqref{Eq:ADPMINMPC} with $\hat{\mathcal{V}}(f(x.w), {\theta})$ inferred from $M$ demonstrations is said to be policy-consistent if, for any $\epsilon >0 $, 
    $\lim_{M\rightarrow \infty} \mathbb{P}(\|\hat{\pi}_\theta(x) - \pi(x)\| >\epsilon) = 0$.
\end{define}
%Policy consistency can be expressed as requiring the residual in optimality conditions to vanish in probability as $M\rightarrow \infty$.

Although the above definition is asymptotic and formal, a natural way to check  consistency in practice is if $\mathcal{D}$ (approximately) satisfies the optimality conditions of \eqref{Eq:ADPMINMPC}. The inverse optimization method of the optimality residual minimization \citep{keshavarz2011imputing} is a promising approach to do this. 
%.  To do this, we employ the inverse optimization method of KKT residual minimization \citep{keshavarz2011imputing}, whereby we exploit the necessary KKT conditions of optimality to learn $\hat{\mathcal{V}}(x,\theta)$ for which a given set of expert demonstrations are approximately \textit{consistent} with \eqref{Eq:ADPMINMPC}. The control policy $\hat{\pi}_\theta:x\rightarrow\mathbb{R}^{n_w}$ obtained by solving \eqref{Eq:ADPMINMPC} is said to be approximately consistent with a set of optimal state-action data pairs $\mathcal{D}=\{(x_i,w_i^*)\}_{i=1}^M$, if $\mathcal{D} $ approximately satisfies the KKT conditions of \eqref{Eq:ADPMINMPC}. Here, the concept of consistency is evaluated qualitatively through control results in Section \ref{sec:benchmarkProblems}. Future work will aim to formally define this concept as it relates to stability under $\hat{\pi}_\theta$ }

The challenge remains that \eqref{Eq:ADPMINMPC} is a mixed-integer OCP, and classical KKT conditions cannot be directly formulated for problems with integer variables, as the required differentiability and constraint qualification conditions fail. Instead, we formulate the KKT conditions of \eqref{Eq:ADPMINMPC} under suitable regularity conditions, but now with $g(\cdot)$ expressed as a continuous relaxation of the integer constraint in \eqref{Eq:ADPMINMPC}, such that $z\in \mathbb{R}^{n_z}$ (i.e., $w = [u,z]^\top \in \mathbb{R}^{n_u+n_z}$): 
%The decision $w$ is approximately optimal for $x$ if residuals of the KKT conditions for the myopic MINMPC \eqref{Eq:ADPMINMPC} with the learned $\hat{\mathcal{V}}(x,\theta)$ are small.
% Here, we expand upon the work of \cite{keshavarz2011imputing}, where we neither require the cost function and constraints of \eqref{Eq:DPMINMPC} to be convex nor the MPC problem to be linear. In doing so, we obtain a generalized method to approximate the parametric function $\hat{\mathcal{V}}(x,\theta)$ for any nonlinear MPC problem with mixed integer decision variables, even if $V_{\eta}(x)$ and $g$ are non-convex.
%The steps to compute $\hat{\mathcal{V}}(x,\theta)$ using KKT residual minimization (c.f. \cite{keshavarz2011imputing} for further detail) are as follows. We first define the KKT conditions of the myopic policy in \eqref{Eq:ADPMINMPC} as 
\begin{subequations}
\begin{align}
	\nabla_w\left( \ell(x,w) + \hat{\mathcal{V}}(f(x,w),\theta)\right) +  \nabla_{w}g(x,w)\lambda  =0,& \\
	g(x,w) \leq 0,&\\
	 \lambda \circ g(x,w)  = 0,&\\
	 \lambda \geq 0.&
\end{align} \label{eq:kkt}%
\end{subequations}
Here, $\circ$ denotes the Hadamard product. For the $i^{\text{th}}$ state-action pair $(x_i,w_i^*)$ in $\mathcal{D}$, we define residuals of the stationarity and complementary slackness  conditions as 
\begin{subequations}
\begin{align}
    r_{\textrm{stat}}^{(i)}(\theta,\lambda_{i}) &=	\nabla_w\left( \ell(x_i,w_i^*) + \hat{\mathcal{V}}(f(x_i,w_i^*),\theta)\right)  \nonumber \\
     & \qquad \ldots +  \nabla_{w}g(x_{i},w_{i}^*)\lambda_i,  \label{eq:aprxstatcond} \\
    r_{\textrm{comp}}^{(i)}(\lambda_{i}) &=\, \lambda_{i}\circ g(x_{i},w_{i}^*). \label{eq:aprxcompcond}
\end{align} \label{eq:kktresiduals}%
\end{subequations}
In line with \cite{abdufattokhov2021learning,keshavarz2011imputing}, we choose a quadratic cost function $\hat{\mathcal{V}}(x,\theta):= x^{\mathsf{T}} \mathbf{P} x :\mathbb{R}^{n_{x}}\rightarrow \mathbb{R}$, parameterized by the matrix $\theta = \mathbf{P}$. 
Collecting the KKT residuals of the dataset $\mathcal{D}$, we can formulate the KKT residual minimization problem as 
\begin{subequations}\label{Eq:PSDP}
	\begin{align}
		\min_{\mathbf{P},\{\lambda_{i}\}}& ~ \sum_{i=1}^M \|r_{\textrm{stat}}^{(i)}(\mathbf{P},\lambda_{i})\|_2^2 + \|r_{\textrm{comp}}^{(i)}(\lambda_{i})\|_2^2,\label{eq:Piocobjective} \\
		 \textrm{s.t.}  \; &\lambda_{i} \geq 0, \; \; i = 1,2,\dots,M, \label{eq:Ppossemidef} \\
    & \mathbf{P} \succ 0, \label{eq:Pposdef}
	\end{align} \label{eq:PIOC}%
\end{subequations} 
where \eqref{eq:Piocobjective} seeks to find a symmetric, positive-definite $\mathbf{P}$ that minimizes the unweighted KKT residuals and \eqref{eq:Ppossemidef} ensures non-negativity of the Lagrange multipliers. Note that the constraint $\mathbf{P} \succ 0$ explicitly excludes the trivial solution $\mathbf{P}=0$ from the feasible set. Once we compute the approximate value function using the relaxed problem, we can then reintroduce the integer constraint when solving \eqref{Eq:ADPMINMPC} online, i.e. $w = [u,z]^\top$, with and $z\in \mathbb{Z}^{n_z}$ as in \eqref{Eq:MINMPC}. 

Although expert demonstrations could originate from a MINMPC, the KKT residual minimization problem \eqref{Eq:PSDP} is formulated for the continuous relaxation of the myopic MINMPC. For such a relaxation, integer actions do not satisfy the stationarity conditions of a continuous optimization problem. The KKT residuals, hence, cannot be driven to zero. Instead, by solving \eqref{Eq:PSDP}, we aim to find the value function $\hat{\mathcal{V}}(x,\theta)$ which minimizes the KKT residuals. The expert demonstrations $(x_{i},w_{i}^*)$ then approximately satisfy the KKT conditions in \eqref{eq:kkt} for $\hat{\mathcal{V}}(x,\theta)$ $i = 1,\ldots,M$ and are, therefore, approximately consistent with the continuous relaxation of \eqref{Eq:ADPMINMPC}. This does not jeopardize the integer constraints in the problem, because the approximate value function is used online in a myopic MINMPC with integer constraints. Thus, the continuous relaxation serves as a differentiable surrogate for learning, while the controller remains fully mixed-integer.

We emphasize that the KKT conditions in~\eqref{eq:kkt} are only necessary optimality conditions for the (generally nonconvex) continuous relaxation of the myopic MPC problem \eqref{Eq:ADPMINMPC}. The learned value function is shaped so that stationary solutions align with that of the observed expert behavior, without a sufficient condition for optimality. This is, however, consistent with how nonlinear MPC problems are practically solved, whereby standard numerical solvers (e.g., SQP or interior-point methods) aim to compute first-order stationary points that satisfy the KKT conditions.  %This is beneficial for MINMPC problems, where $V_{\eta}(x)$ could be piecewise, non-smooth, and non-convex. Approximating $V_{\eta}(x)$ arbitrarily though minimization of mean-squared error or other similar loss metrics \citep{ab2020inverse, akhtar2021learning} may not be possible for such a case. Instead, by exploiting the KKT conditions in this manner, we may yet find a cost-to-go representation $\hat{\mathcal{V}}(x,\theta)$  for which the state-action pairs in $\mathcal{D}$ remain consistent with \eqref{Eq:ADPMINMPC}. The result is then myopic MINMPC \eqref{Eq:ADPMINMPC} that approximates the optimal actions ${w^*}(x)$ of \eqref{Eq:MINMPC}. 
%Still, we must choose the class of $\hat{\mathcal{V}}(x,\theta)$. %\cite{keshavarz2011imputing} propose using 
%One option, in line with \cite{keshavarz2011imputing}, is to use a quadratic cost function $\hat{\mathcal{V}}(x,\theta):= x^{\mathsf{T}} \mathbf{P} x :\mathbb{R}^{n_{x}}\rightarrow \mathbb{R}$, parameterized by the matrix $\theta = \mathbf{P}$
%where the requirement \eqref{eq:Pposdef} that $\mathbf{P}$ be symmetric, positive-definite ensures that $\hat{\mathcal{V}}$ is a positive-definite quadratic form and thus a reasonable candidate Lyapunov function. In practice, this structural choice promotes (but does not strictly enforce) a descent-like behavior as in \eqref{eq:descentcons}.

% For \eqref{eq:Piocobjective}, we substitute $\mathbf{P}$ for $\theta$ in \eqref{eq:aprxstatcond} and expand with the chain rule, yielding
% \begin{align}
%     r_{\textrm{stat}}^{(i)}(\mathbf{P},\lambda_{i})
% & = \nabla_w \ell(x_i,w_i^*)
% + 2 \nabla_w f(x_i,w_i^*)^{\mathsf{T}} \mathbf{P} f(x_i,w_i^*) \nonumber \\
% & \qquad \ldots + \nabla_{w}g(x_{i},w_{i}^*)^{\mathsf{T}} \lambda_{i}
% \end{align}
Lastly, by deliberately restricting the hypothesis class to a quadratic function, the KKT residual minimization problem becomes an SDP. Consistent with VC-dimension theory \citep{vidyasagar2003vapnik}, this reduces the effective complexity of learning a value function and improves sample efficiency, enabling recovery of sufficiently-accurate functions from only a small number of demonstrations. This is particularly important given the high cost of generating the offline samples through MINLP solvers. However, there are other choices for $\hat{\mathcal{V}}(\cdot)$. For instance, one could attempt to learn $\theta$ for an arbitrary function approximator, shaped by enforcing a descent condition as done in \cite{DK2025c2gcdc}. This more generalized approach may improve performance, but comes at the expense of requiring a larger training dataset and possibly non-convex optimization depending on the choice of $\hat{\mathcal{V}}(\cdot)$. Yet, enforcing a descent condition on the learning of $\hat{\mathcal{V}}(\cdot)$ may improve stability under the resulting myopic MINMPC policy.

\section{Illustrative examples}
\label{sec:benchmarkProblems}

We apply the proposed method to two systems: the Lotka-Volterra fishing problem and the control of a satellite in orbit.  For each system, the MINMPC takes the form \eqref{Eq:MINMPC} with objective function $\ell(x,w) := x^{\mathsf{T}}Qx + w^{\mathsf{T}}Rw$, tuning parameters $Q$ and $R$, and transcribed with direct multiple shooting using 4th-order Runge-Kutta discretization %with sampling time $T_s$ is applied to both systems, yielding $x_{k+1} = f(x_k,w_k)$. 
We construct the controllers in \texttt{Matlab} with the \texttt{CasADi} toolbox \citep{Andersson2018CasADi}. All MINLPs are warm-started and solved using  \texttt{Bonmin} \citep{bonami2008algorithmic}. We generate offline data for each system using the original MINMPCs. We aim to learn $x^{\mathsf{T}} \mathbf{P} x$ of the myopic MINMPC with $ \eta = 1 $. Each SDP \eqref{Eq:PSDP} is formulated with the \texttt{YALMIP} toolbox and computed using the \texttt{MOSEK} SDP solver \citep{Lofberg2004, mosek}. The systems here have states and inputs $\mathcal{O}(1)$, thus no normalization is applied (though normalization may be desirable for multi-magnitude systems). We compare the myopic MINMPC against the original MINMPC in terms of reference tracking performance and computation cost in discrete-time simulations with added plant-model mismatch. All computations were performed on an Intel(R) Core\textsuperscript{TM} i5 processor running at 2.40 GHz with 15.7 GB of usable RAM.

\subsection{Lotka-Volterra fishing problem} We first consider the Lotka-Volterra fishing problem \citep{sager2006numerical}, a foundational benchmark example for mixed-integer optimal control problem. This problem applies an optimal fishing allowance to stabilize predator and prey populations. The dynamic system \eqref{eq:fishingproblem} describes the coupled evolution of the prey population $x_1$ and the predator population $x_2$ in continuous time as
\begin{subequations}
\begin{gather}
 \begin{bmatrix} \dot{x}_1  \\ \dot{x}_2 \end{bmatrix}
 =
  \begin{bmatrix}
   x_1 - x_1x_2 - c_1x_1z \\
   -x_2 + x_1x_2 - c_2x_2z
   \end{bmatrix},
   \label{eq:fishingdynamics} \\
   z \in \{0,1\}, \label{eq:binary} \\
   [x_1, x_2]^\top \geq 0. \label{eq:fishcons}
\end{gather}
\label{eq:fishingproblem}%
\end{subequations}
The binary decision variable $w = z$ represents the decision to fish ($z=1$) or not to fish ($z=0$) with rate coefficients $c_1 = 0.4$ and $c_2 = 0.2$. The continuous-time system is discretized with $T_s = 0.1$ (s). For the MINMPC formulation, $Q = \text{diag}([1, 1])$, $R = 0.01$, $N = 30$. 

\subsubsection{Learning the cost-to-go offline} 
\begin{figure}[t!]
\centering
\includegraphics[width=0.92\columnwidth]{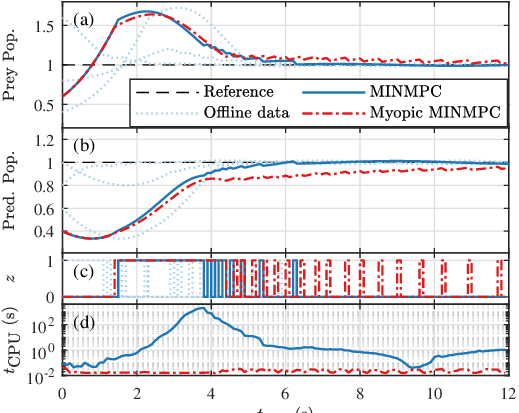}
\caption{(a) Prey and (b) predator state populations for the expert demonstrations (in light blue), MINMPC controller (in blue), and myopic MINMPC controller (in red). The (c) control decisions and (d) the computation time per controller decision.}
\label{fig:fishing}%
\end{figure}

To compute $ \mathbf{P} $, the original MINMPC was run in an offline feedback control simulation with perfect state observation for three initial states, resulting a data set $\mathcal{D}$ of size $M=360$ consisting of three trajectories. Note that $\mathcal{D}$ in this case was generated \textit{without} relaxing the binary constraint \eqref{eq:binary}. These offline expert demonstrations are shown in light blue dotted lines in Fig. \ref{fig:fishing}. The dataset required $t_{\textrm{CPU}}=8.42\times10^3$ ({s}) in total to solve. We solved the SDP \eqref{eq:PIOC} for the $\mathbf{P}$ given the dataset $\mathcal{D}$, yielding
\begin{gather*}
 \mathbf{P} = \begin{bmatrix}
    0.5965  &  0.4627 \\
    0.4627  &  0.3589 \end{bmatrix},
   \label{eq:fishPmatrix}%
\end{gather*}
 The maximum KKT residuals, $\|r_{stat}\|_\infty = 9.70\times10^{-5}$, and $\|r_{comp}\|_\infty = 9.67\times10^{-5}$ are small compared to the magnitude of $\mathbf{P}$. This suggests that \eqref{Eq:ADPMINMPC} using $\hat{\mathcal{V}}(x,\theta):= x^{\mathsf{T}} \mathbf{P} x$ is approximately consistent with the expert demonstrations.

\subsubsection{Online controller performance}
The controllers were designed to track a reference $[x_1^{\text{ref}}, x_2^{\text{ref}}]^{\mathsf{T}} = 1$. Plant-model mismatch was implemented as $10\%$ error in the fishing coefficients, with the true $c_1 = 0.44$ and $c_2 = 0.22$ in the plant simulator. The results are given in Fig. \ref{fig:fishing}. As shown in Fig. \ref{fig:fishing}a and \ref{fig:fishing}b, the myopic MINMPC controller produces a similar state trajectory to that of the long prediction horizon MINMPC controller, but at a significantly lower CPU time as shown in \ref{fig:fishing}d. Note that a MINMPC controller with a short prediction horizon ($\ll N$) and without the learned cost-to-go does not drive the state to the reference.

\subsection{Satellite problem}
We consider the control of a simple satellite of unit mass. The satellite must maintain a desired altitude using impulsive thrust in the radial $\hat{e}_r$ and angular $\hat{e}_{\theta}$ directions. The satellite's orbital dynamics are defined in the polar coordinate frame in continuous time as
\begin{subequations}
\begin{gather}
 \begin{bmatrix} \dot{x}_1  \\ \dot{x}_2 \\ \dot{x}_3 \end{bmatrix}
 =
  \begin{bmatrix}
   x_2 \\
   x_1 x_3^2 - \frac{d_1}{x_1^2} + d_3 z_1 \\
   - \frac{2 x_2 x_3}{x_1} -  \frac{d_2 x_3^2}{x_1} + \frac{d_3 z_2}{x_1}
   \end{bmatrix},
   \label{eq:satellitedynamics} \\
   z_1, z_2 \in \{-1,0,1\}, \label{eq:combinatorial} \\
   3 \leq x_1 \leq 7, \label{eq:satcons}
\end{gather}
\label{eq:satelliteproblem}%
\end{subequations}
where the celestial body about which the satellite orbits is the origin. The states are the radius $r = x_1$, the radial velocity $\dot{r} = x_2$, and the angular velocity $\dot{\theta} = x_3$. The thrust in $\hat{e}_r$ is $z_1$, the thrust in $\hat{e}_{\theta}$ is $z_2$, $d_1 = 2$ is the gravitational constant, $d_2 = 0.1$ is atmospheric drag, $d_3 = 0.1$ is the thrust efficiency, and the constraints on $x_1$ represent the \enquote{safe} orbital region of the satellite. The continuous-time system is discretized with $T_s = 0.5$ (s). The MINMPC $Q = \text{diag}([10, 1, 1])$ and $R = \text{diag}([1, 1])$, and the prediction horizon is $N=30$. 

\subsubsection{Learning the cost-to-go offline}
The computational complexity of the satellite problem is significantly greater than that of the Lotka-Volterra problem. The Lotka-Volterra problem has a single binary decision variable, meaning computation time necessary to solve the underlying MINLP at each time step scales $\propto 2^N$. However, the satellite problem has two decision variables with three distinct integer values each, meaning computation time scales $\propto 3^{2N}$. Generating even a single satellite trajectory required $t_{\textrm{CPU}}=1.26 \times 10^4$ (s) to compute.

\begin{figure}[t!]
\centering
\includegraphics[width=0.92\columnwidth]{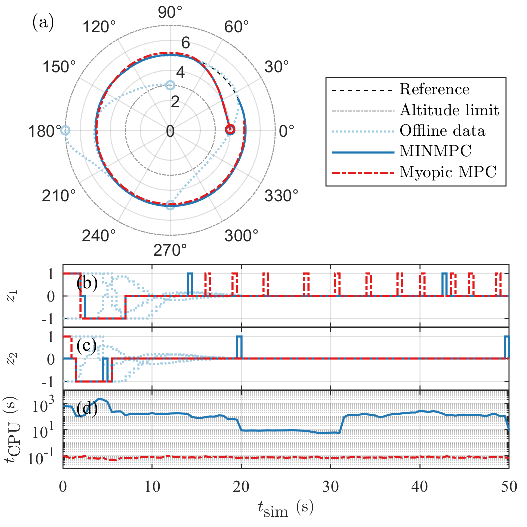}
\caption{(a) Polar plot of the satellite trajectories for the offline expert demonstrations (light blue), MINMPC (blue), and myopic MINMPC (red). Each initial state is denoted by a circle. (b) $z_1$ and (c) $z_2$  for each controller. (d) computation time per controller decision.}
\label{fig:satellite}
\end{figure}

This computational cost motivated generation of training data for cost-to-go learning using a relaxed formulation of the satellite problem, where $z_1, z_2 \in [-1,1]$ rather than $z_1, z_2 \in \{-1,0,1\}$. The integer constraint were then only introduced for the online myopic MINMPC. To this end, we generated three trajectories without integer constraints using a relaxed NMPC controller offline, solved with \texttt{IPOPT} \citep{wachter2006ipopt}. The resulting dataset of size $M=120$ required only $t_{\textrm{CPU}}=2.06$ ({s}) to solve in total. The offline  expert demonstrations are shown as light blue dotted-lines in Fig. \ref{fig:satellite}. We then solved for $\mathbf{P}$ in \eqref{eq:PIOC} given the dataset of NMPC trajectories, yielding  \begin{gather*}
 \mathbf{P} = \begin{bmatrix} 48.7 &  81.4 &  81.5 \\ 81.4  & 164  & 239 \\
 81.5 & 239 & 509 \\ \end{bmatrix},
   \label{eq:satPmatrix}
\end{gather*}
with $\|r_{stat}\|_\infty = 0.663$ and $\|r_{comp}\|_\infty = 0.296$. As with the fishing problem, the residuals are small relative to the magnitude of $\mathbf{P}$, suggesting approximate policy consistency. 

\subsubsection{Online controller performance}
We examine the simulation results for the satellite problem using the original MINMPC and the myopic MINMPC in Fig. \ref{fig:satellite}. The controllers tracked a reference $[x_1^{\text{ref}}, x_2^{\text{ref}}, x_3^{\text{ref}}]^{\mathsf{T}} = [5,0,0.126]^{\mathsf{T}}$. Plant-model mismatch was introduced as variable drag $d_3(x_1) = \exp{((5-x_1+2\log(0.1))/2)}$ in the plant simulator. Note that, although the training data did not consider integer decision variables, we compare the performance of the myopic MINMPC with the full horizon MINMPC, both of which consider integer decision variables. Despite learning the cost-to-go from NMPC demonstrations without integer constraints, the myopic MINMPC produces a similar trajectory to the original MINMPC. Without the learned cost-to-go, a MINMPC with a short prediction horizon drives the satellite into the upper state constraint.

\subsection{Computational performance}
\begin{table}[htbp]
    \centering
    \label{tab:tcpu}
    \begin{tabular}{cccc}
        \toprule
        Benchmark & Total offline &\multicolumn{2}{c}{Maximum online $t_{\text{CPU}}$ (s) }   \\
        problem & $t_{\text{CPU}}$ for $\mathcal{D}$ (s) & MINMPC &myopic MINMPC \\
        \midrule
        Fishing & $8.42\times10^3$ & $2.03\times10^3$ & $4.62\times10^{-2}$ \\
        Satellite & 2.06 & $2.16\times10^3$ & $1.01\times10^{-1}$ \\
        \bottomrule
    \end{tabular}
\end{table}
 
We compare the computational performance of each method, shown in Fig. \ref{fig:fishing}d and Fig. \ref{fig:satellite}d, in the table below. Both of the original MINMPC controllers were very computationally expensive, each exhibiting maximum computation times of over 30 min per solution. However, the maximum computation times of the myopic MINMPC controllers were only $4.62\times10^{-2}$ ({s}) and $1.01\times10^{-1}$ ({s}) for the fishing and satellite problems, respectively. The myopic MINMPC is consistently over four orders of magnitude faster than the original MINMPC, demonstrating the potential of this approach

\section{Conclusion}
\label{sec:conclusion}

In this work, we have shown that the KKT residual minimization method could be a promising means to approximate the cost-to-go for a myopic MINMPC from a set of expert demonstrations in the form of optimal state-action pairs computed offline. We show that, for two illustrative examples, the myopic MINMPC controller using the learned cost-to-go closely replicates the reference tracking performance of the original, long prediction horizon MINMPC controller, but at a significant reduction in computation time per controller decision. While we approximate the cost-to-go with a quadratic function because the resulting OCP is easily solvable as an SDP, a quadratic term may not be the best approach for a given MINMPC problem. Other approximate cost-to-go function classes, including general function approximators, may improve the optimality residual. While residuals are evaluated at finite samples, future work involves generalization to unseen samples and the impact of this on closed-loop performance. %Future work expanding on that of \cite{DK2025c2gcdc} is exploring whether enforcing a descent property in the learning process may prove stability through the scenario approach. However, recursive feasibility may be more difficult to prove in this setting due to the computational cost of recomputing terminal sets online.  

\bibliography{BibL4DC}             % bib file to produce the bibliography

\end{document}